\documentclass{article}
\usepackage{amssymb}

\usepackage{amsmath}
\usepackage{graphicx}

\newtheorem{theorem}{Theorem}[section]

\newtheorem{definition}[theorem]{Definition}
\newtheorem{example}[theorem]{Example}

\newtheorem{notation}[theorem]{Notation}

\newtheorem{remark}[theorem]{Remark}

\input{tcilatex}

\begin{document}

\title{Defining the Delays of the Asynchronous Circuits}
\author{Serban. E. Vlad \\
Oradea City Hall, Piata Unirii, Nr. 1, 3700, Oradea, Romania\\
serban\_e\_vlad@yahoo.com, http://site.voila.fr/serban\_e\_vlad}
\date{}
\maketitle

\begin{abstract}
The purpose of the paper is that of defining the delays of a circuit as well
as the properties of: determinism, order, time invariance, constancy,
symmetry and the serial connection.
\end{abstract}

\section{Introduction}

Digital electrical engineering is a non-formalized theory and the aim of our
concerns is that of trying a semi-formalization. The delay (condition) is
the proposed starting point and it represents the real time model of the
circuit that computes the identity function $1_{\{0,1\}}$. Logical gates and
wires are modeled by a Boolean function that computes instantaneously, in
real time, the output depending on the inputs and by zero, one or several
delays at the output or at the inputs. The model of an asynchronous circuit
consists then in the composition of the models of the logical gates and
wires, meaning the serial connection of the delays and the composition of
the Boolean functions.

\section{Preliminaries}

\begin{definition}
$\mathbf{B}=\{0,1\}$ is endowed with the discrete topology, with the order $%
0\leq1$ and with the usual laws: $^{-},\cdot,\cup,\oplus$.
\end{definition}

\begin{definition}
Let $x:\mathbf{R}\rightarrow\mathbf{B}$ and $A\subset\mathbf{R}$. We define%
\begin{align*}
\underset{\xi\in A}{\bigcap}x(\xi) & =\left\{ 
\begin{array}{c}
0,\exists\xi\in A,x(\xi)=0 \\ 
1,otherwise\ \ \ \ \ \ \ \ 
\end{array}
\right. \quad\underset{\xi\in\varnothing}{\bigcap}x(\xi)=1 \\
\underset{\xi\in A}{\bigcup}x(\xi) & =\left\{ 
\begin{array}{c}
1,\exists\xi\in A,x(\xi)=1 \\ 
0,otherwise\ \ \ \ \ \ \ 
\end{array}
\right. \quad\underset{\xi\in\varnothing}{\bigcup}x(\xi)=0
\end{align*}
\end{definition}

\begin{definition}
The order and the laws of $\mathbf{B}$ induce an order and laws in the set
of the $\mathbf{R}\rightarrow\mathbf{B}$ functions, that are noted with the
same symbols.
\end{definition}

\begin{definition}
Let $x:\mathbf{R}\rightarrow\mathbf{B}$. The \textit{left limit} function $%
x(t-0)$ is defined by%
\begin{equation*}
\forall t\in\mathbf{R},\exists\varepsilon>0,\forall\xi\in(t-\varepsilon
,t),x(\xi)=x(t-0)
\end{equation*}
\end{definition}

\begin{definition}
We suppose that $x(t-0)$ exists. Then the functions $\overline{x(t-0)}\cdot
x(t)$, $x(t-0)\cdot\overline{x(t)}$ are called the \textit{left
semi-derivatives} of $x$.
\end{definition}

\begin{definition}
The \textit{characteristic function} $\chi_{A}:\mathbf{R}\rightarrow \mathbf{%
B}$ of the set $A\subset\mathbf{R}$ is%
\begin{equation*}
\chi_{A}(t)=\left\{ 
\begin{array}{c}
1,t\in A \\ 
0,t\notin A%
\end{array}
\right.
\end{equation*}
\end{definition}

\begin{definition}
We call \textit{signal} a function $x$ having the property that the
unbounded sequence $0\leq t_{0}<t_{1}<t_{2}<...$ exists so that%
\begin{equation*}
x(t)=x(t_{0}-1)\cdot\chi_{(-\infty,t_{0})}(t)\oplus
x(t_{0})\cdot\chi_{\lbrack t_{0},t_{1})}(t)\oplus x(t_{1})\cdot\chi_{\lbrack
t_{1},t_{2})}(t)\oplus...
\end{equation*}
and we note with $S$ the set of the signals.
\end{definition}

\begin{notation}
$\tau^{d}:\mathbf{R}\rightarrow\mathbf{R}$ is the translation $\tau
^{d}(t)=t-d$, where $t,d\in\mathbf{R}$.
\end{notation}

\begin{theorem}
The constant functions $0,1:\mathbf{R}\rightarrow\mathbf{B}$ are signals. If 
$\ 0\leq m\leq d$ and $x,y\in S$, then the functions $x\circ\tau^{d},%
\overline{x(t)},x(t)\cdot y(t),x(t)\cup y(t),x(t)\oplus y(t),\underset{\xi
\in\lbrack t-d,t-d+m]}{\bigcap}x\left( \xi\right) ,$\ $\underset{\xi
\in\lbrack t-d,t-d+m]}{\bigcup}x\left( \xi\right) $ are signals too.
\end{theorem}

\begin{theorem}
$\forall x\in S,$ the left limit function $x(t-0)$ exists.
\end{theorem}

\begin{notation}
We note with $P^{\ast}(S)$ the set of the non-empty subsets of $S$.
\end{notation}

\section{Stability. Rising and Falling Transmission Delays for Transitions}

\begin{definition}
Let $u,x\in S$, called \textit{input} and respectively \textit{state} (or 
\textit{output}). The implication%
\begin{equation*}
\forall a\in \mathbf{B},(\exists t_{1},\forall t\geq
t_{1},u(t)=a)\Longrightarrow (\exists t_{2},\forall t\geq t_{2},x(t)=a)
\end{equation*}%
is called the \textit{stability condition} (SC). We say that the couple $%
(u,x)$ satisfies SC. We call also SC the function $Sol_{SC}:S\rightarrow
P^{\ast }(S)$ defined by%
\begin{equation*}
Sol_{SC}(u)=\{x|(u,x)\ satisfies\ SC\}
\end{equation*}
\end{definition}

\begin{definition}
We suppose the existence of $a\in\mathbf{B}$ so that $\exists t_{1},\forall
t\geq t_{1},u(t)=a$ and the fact that $(u,x)$ satisfies SC. If $u,x$ are
both non-constant, we note%
\begin{equation*}
t_{1}^{\ast}=\min\{t_{1}|\forall t\geq t_{1},u(t)=a\},\
t_{2}^{\ast}=\min\{t_{2}|\forall t\geq t_{2},x(t)=a\}
\end{equation*}
The \textit{transmission delay for transitions} is the number $d\geq0$
defined by%
\begin{equation*}
d=\max(0,t_{2}^{\ast}-t_{1}^{\ast})
\end{equation*}
If $\overline{u(t_{1}^{\ast}-0)}\cdot u(t_{1}^{\ast})=\overline{%
x(t_{2}^{\ast }-0)}\cdot x(t_{2}^{\ast})=1$, then $d$ is called \textit{%
rising} and if $u(t_{1}^{\ast}-0)\cdot\overline{u(t_{1}^{\ast})}%
=x(t_{2}^{\ast}-0)\cdot\overline{x(t_{2}^{\ast})}=1$, then $d$ is called 
\textit{falling}. If $u$, respectively $x$ is constant, then $t_{1}^{\ast}$
respectively $t_{2}^{\ast}$ is by definition $0$.
\end{definition}

\section{Delays}

\begin{definition}
\label{d41}A \textit{delay condition} (DC) or shortly a \textit{delay} is a
function $i:S\rightarrow P^{\ast}(S)$ with the property that $\forall
u,i(u)\subset Sol_{SC}(u)$.
\end{definition}

\begin{remark}
The problem of the delays is that of the real time computation of the
identity function $1_{\mathbf{B}}$. In practice we often work with systems
of equations and inequalities in $u,x$ that model this computation and $i(u)$
represents for all $u$ the set of the solutions of these systems. Definition %
\ref{d41} requests that solutions exist for any $u$ and that the systems be
stable.
\end{remark}

\begin{example}
\label{p43} The next functions are DC's:

\begin{itemize}
\item $i(u)=\{u\}$ is usually noted with $I$. More general, the equation $%
i(u)=\{u\circ\tau^{d}\}$ defines a DC noted with $I_{d},d\geq0$.

\item $i(u)=\{x|\exists d\geq0,x(t)=u(t)\cdot\chi_{\lbrack d,\infty)}(t)\}$

\item $i(u)=Sol_{SC}(u)$
\end{itemize}
\end{example}

\begin{theorem}
Let $U\subset S$ and the DC's $i,j$.
\end{theorem}

\begin{description}
\item[a)] If $\forall u,i(u)\wedge U\neq\emptyset$, then the next equation
defines a DC%
\begin{equation*}
(i\wedge U)(u)=i(u)\wedge U
\end{equation*}

\item[b)] If $i,j$ satisfy $\forall u,i(u)\wedge j(u)\neq\emptyset$, then $%
i\wedge j$ is a DC defined by%
\begin{equation*}
(i\wedge j)(u)=i(u)\wedge j(u)
\end{equation*}

\item[c)] Items a), b) are generalized by taking an arbitrary function $%
\varphi:S\rightarrow P^{\ast}(S)$ with $\forall u,i(u)\wedge\varphi
(u)\neq\emptyset$; $i\wedge\varphi$ is a DC%
\begin{equation*}
(i\wedge\varphi)(u)=i(u)\wedge\varphi(u)
\end{equation*}

\item[d)] $i$ and $j$ define the DC $i\vee j$ in the next manner:%
\begin{equation*}
(i\vee j)(u)=i(u)\vee j(u)
\end{equation*}
\end{description}

\section{Determinism}

\begin{definition}
The DC $i$ is called \textit{deterministic} if $\forall u,i(u)$ has a single
element and \textit{non-deterministic} otherwise.
\end{definition}

\begin{remark}
By interpreting $i$ as the set of the solutions of a system, its determinism
indicates the uniqueness of the solution for all $u$. On the other hand we
shall identify the deterministic DC's with the functions $i:S\rightarrow S$.
The non-deterministic delays are justified by the fact that in an electrical
circuit to one input $u$ there corespond several possible outputs $x$
depending on the variations in ambient temperature, power supply, on the
technology etc.
\end{remark}

\begin{example}
In \ref{p43} $I,I_{d}$ are deterministic and the other delays are
non-deterministic. Let $U\subset S$ and the DC's $i,j$ with $i$
deterministic. If $\forall u,i(u)\wedge U\neq\emptyset$, then $i\wedge U(=i)$
is deterministic and similarly for $i\wedge j$.
\end{example}

\section{The Order}

\begin{definition}
For the DC's $i,j$ we define%
\begin{equation*}
i\subset j\Longleftrightarrow\forall u,i(u)\subset j(u)
\end{equation*}
\end{definition}

\begin{remark}
The inclusion $\subset$ defines an order in the set of the DC's. $Sol_{SC}$
is the universal element relative to this order, because any $i$ satisfies $%
i\subset Sol_{SC}$. We interpret the inclusion $i\subset j$ by the fact that
the first system contains more restrictive conditions than the second and
the model in the first case is more precise than in the second one. In
particular, a deterministic DC contains the maximal information and the DC $%
Sol_{SC}$ contains the minimal information about the modeled circuit.
\end{remark}

\begin{theorem}
Any DC $j$ includes a deterministic DC $i$; if $i\subset j$ and if $j$ is
deterministic, then $i=j$.
\end{theorem}

\section{Time Invariance}

\begin{definition}
The DC $i$ is called \textit{time invariant} if%
\begin{equation*}
\forall u,\forall x,\forall d\in\mathbf{R},(u\circ\tau^{d}\in S\ and\ x\in
i(u))\Longrightarrow(x\circ\tau^{d}\in S\ and\ x\circ\tau^{d}\in i(u\circ
\tau^{d}))
\end{equation*}
and if the previous property is not satisfied then $i$ is called \textit{%
time variable}.
\end{definition}

\begin{example}
$I_{d}$ is time invariant, $d\geq0$. Let the time invariant DC's $i,j$ with $%
\forall u,i(u)\wedge j(u)\neq\emptyset$; then $i\wedge j$ is time invariant.
Let $k$ time invariant; then $i\vee k$ is time invariant. $Sol_{SC}$ is time
variable.
\end{example}

\begin{theorem}
If $i$ is a time invariant DC, then the next equivalence holds:%
\begin{equation*}
\forall u,\forall x,\forall d\geq0,x\in i(u)\Longleftrightarrow x\circ\tau
^{d}\in i(u\circ\tau^{d})
\end{equation*}
\end{theorem}

\section{Constancy}

\begin{definition}
A DC $i$ is called \textit{constant} if $\exists d_{r}\geq0,\exists
d_{f}\geq0$ so that $\forall u,\forall x\in i(u)$ we have%
\begin{align*}
\overline{x(t-0)}\cdot x(t) & \leq u(t-d_{r}) \\
x(t-0)\cdot\overline{x(t)} & \leq\overline{u(t-d_{f})}
\end{align*}
If the previous property is not satisfied, then $i$ is called \textit{%
non-constant}.
\end{definition}

\begin{example}
$I_{d}$ is constant, $d\geq0$. Let $U\subset S$ and the DC's $i,j$, the
first constant. If $i\wedge U$ and $i\wedge j$ are defined, then they are
constant. More general, any DC included in a constant DC is constant.
\end{example}

\begin{theorem}
\label{t83}The next functions%
\begin{equation*}
x(t)=\underset{\xi \in \lbrack t-d,t-d+m]}{\bigcap u\left( \xi \right) }%
,\quad x(t)=\underset{\xi \in \lbrack t-d,t-d+m]}{\bigcup u\left( \xi
\right) }
\end{equation*}%
are deterministic, time invariant, constant DC's, where $0\leq m\leq d$.
\end{theorem}

\begin{remark}
Constancy means that $x$ is allowed to switch only if $u$ has anticipated
this possibility $d_{r}$, respectively $d_{f}$ time units before. Its
satisfaction does not imply the uniqueness of $d_{r},d_{f}$ and \ref{t83}
offers such a counterexample.
\end{remark}

\section{Rising-Falling Symmetry}

\begin{definition}
The DC $i$ is called (\textit{rising-falling}) \textit{symmetrical} if%
\begin{equation*}
\forall u,i(\overline{u})=\{\overline{x}|x\in i(u)\}
\end{equation*}
and respectively (\textit{rising-falling}) a\textit{symmetrical} otherwise.
\end{definition}

\begin{example}
$I_{d},d\geq0$ and $Sol_{SC}$ are symmetrical. Let the symmetrical DC's $i,j 
$; if $i\wedge j$ is defined, then it is symmetrical. The DC $i\vee j$ is
symmetrical too.
\end{example}

\section{The Serial Connection}

\begin{definition}
For the DC's $i,j$ we note with $k=i\circ j$ the function $k:S\rightarrow
P^{\ast}(S)$ defined by%
\begin{equation*}
k(u)=\{y|\exists x,x\in j(u)\ and\ y\in i(x)\}
\end{equation*}
$k$ is called the \textit{serial connection} of the DC's $i,j$.
\end{definition}

\begin{theorem}
The next statements are true:

\begin{description}
\item[a)] $k$ is a DC.

\item[b)] If $i,j$ are deterministic, then $k$ is deterministic.

\item[c)] If $i,j$ are time invariant, then $k$ is time invariant.

\item[d)] If $i,j$ are symmetrical, then $k$ is symmetrical.
\end{description}
\end{theorem}

\begin{remark}
The serial connection of the constant delays is not constant, in general.
The set of the DC's is a non-commutative semi-group relative to the serial
connection and $I$ is the unit.
\end{remark}

\begin{theorem}
Let the DC's $i,j,k$. The next implications are true:%
\begin{align*}
i & \subset j\Longrightarrow i\circ k\subset j\circ k \\
j & \subset k\Longrightarrow i\circ j\subset i\circ k
\end{align*}
\end{theorem}

\begin{theorem}
Let $U\subset S$ and the DC's $i,j,k$.
\end{theorem}

\begin{description}
\item[a)] If $\forall u,i(u)\wedge U\neq\emptyset$, then $\forall u,(i\circ
j)(u)\wedge U\neq\emptyset$ and%
\begin{equation*}
(i\wedge U)\circ j=(i\circ j)\wedge U
\end{equation*}
If $\forall u,j(u)\wedge U\neq\emptyset$, then we have%
\begin{equation*}
i\circ(j\wedge U)\subset i\circ j
\end{equation*}

\item[b)] If $\forall u,i(u)\wedge j(u)\neq\emptyset$, then $\forall
u,(i\circ k)(u)\wedge(j\circ k)(u)\neq\emptyset$ and%
\begin{equation*}
(i\wedge j)\circ k\subset(i\circ k)\wedge(j\circ k)
\end{equation*}
If $\forall u,j(u)\wedge k(u)\neq\emptyset$, then $\forall u,(i\circ
j)(u)\wedge(i\circ k)(u)\neq\emptyset$ and%
\begin{equation*}
i\circ(j\wedge k)\subset(i\circ j)\wedge(i\circ k)
\end{equation*}

\item[c)] We have%
\begin{align*}
(i\vee j)\circ k & =(i\circ k)\vee(j\circ k) \\
i\circ(j\vee k) & =(i\circ j)\vee(i\circ k)
\end{align*}
\end{description}

\end{document}